\documentclass[a4paper, oneside, twocolumn, notitlepage, 10pt]{extarticle_ecoc}
\usepackage[nolist]{acronym}
\usepackage{tikz}
\usepackage{pgfplots}
\usetikzlibrary{shapes,arrows,positioning,calc, matrix,spy,patterns}
\usetikzlibrary{decorations.pathreplacing,calligraphy}
\usetikzlibrary{colorbrewer}
\usetikzlibrary{fit}
\usepgfplotslibrary{colorbrewer}
\usepackage{subcaption}
\usepackage{booktabs}
\usepackage{ecoc}

\usetikzlibrary{external}
\definecolor{kit-green100}{rgb}{0,.59,.51}
\definecolor{kit-green70}{rgb}{.3,.71,.65}
\definecolor{kit-green50}{rgb}{.50,.79,.75}
\definecolor{kit-green30}{rgb}{.69,.87,.85}
\definecolor{kit-green15}{rgb}{.85,.93,.93}
\definecolor{KITgreen}{rgb}{0,.59,.51}

\definecolor{KITpalegreen}{RGB}{130,190,60}
\colorlet{kit-maigreen100}{KITpalegreen}
\colorlet{kit-maigreen70}{KITpalegreen!70}
\colorlet{kit-maigreen50}{KITpalegreen!50}
\colorlet{kit-maigreen30}{KITpalegreen!30}
\colorlet{kit-maigreen15}{KITpalegreen!15}

\definecolor{KITblue}{rgb}{.27,.39,.66}
\definecolor{kit-blue100}{rgb}{.27,.39,.67}
\definecolor{kit-blue70}{rgb}{.49,.57,.76}
\definecolor{kit-blue50}{rgb}{.64,.69,.83}
\definecolor{kit-blue30}{rgb}{.78,.82,.9}
\definecolor{kit-blue15}{rgb}{.89,.91,.95}

\definecolor{KITyellow}{rgb}{.98,.89,0}
\definecolor{kit-yellow100}{cmyk}{0,.05,1,0}
\definecolor{kit-yellow70}{cmyk}{0,.035,.7,0}
\definecolor{kit-yellow50}{cmyk}{0,.025,.5,0}
\definecolor{kit-yellow30}{cmyk}{0,.015,.3,0}
\definecolor{kit-yellow15}{cmyk}{0,.0075,.15,0}

\definecolor{KITorange}{rgb}{.87,.60,.10}
\definecolor{kit-orange100}{cmyk}{0,.45,1,0}
\definecolor{kit-orange70}{cmyk}{0,.315,.7,0}
\definecolor{kit-orange50}{cmyk}{0,.225,.5,0}
\definecolor{kit-orange30}{cmyk}{0,.135,.3,0}
\definecolor{kit-orange15}{cmyk}{0,.0675,.15,0}

\definecolor{KITred}{rgb}{.63,.13,.13}
\definecolor{kit-red100}{cmyk}{.25,1,1,0}
\definecolor{kit-red70}{cmyk}{.175,.7,.7,0}
\definecolor{kit-red50}{cmyk}{.125,.5,.5,0}
\definecolor{kit-red30}{cmyk}{.075,.3,.3,0}
\definecolor{kit-red15}{cmyk}{.0375,.15,.15,0}

\definecolor{KITpurple}{RGB}{160,0,120}
\colorlet{kit-purple100}{KITpurple}
\colorlet{kit-purple70}{KITpurple!70}
\colorlet{kit-purple50}{KITpurple!50}
\colorlet{kit-purple30}{KITpurple!30}
\colorlet{kit-purple15}{KITpurple!15}

\definecolor{KITcyanblue}{RGB}{80,170,230}
\colorlet{kit-cyanblue100}{KITcyanblue}
\colorlet{kit-cyanblue70}{KITcyanblue!70}
\colorlet{kit-cyanblue50}{KITcyanblue!50}
\colorlet{kit-cyanblue30}{KITcyanblue!30}
\colorlet{kit-cyanblue15}{KITcyanblue!15}

\definecolor{KITbraun}{RGB}{167,130,46}

\usepackage{colortbl}
\usepackage{xcolor}
\usepackage{pgfmath} %

\begin{document}
\selectlanguage{english}    %

\begin{acronym}[TROLL]
    \acro{ASE}[ASE]{amplified spontaneous emission}
	\acro{AWGN}[AWGN]{additive white Gaussian noise}
    \acro{CD}[CD]{chromatic dispersion}
    \acro{CDC}[CDC]{chromatic dispersion compensation}
    \acro{CPR}[CPR]{carrier phase recovery}
    \acro{DSP}[DSP]{digital signal processing}
	\acro{EEPN}[EEPN]{equalization-enhanced phase noise}
    \acro{FDPE}[FDPE]{frequency-dependent phase error}
    \acro{GN}[GN]{Gaussian noise}
    \acro{IDR}[IDR]{ideal data remodulation}
	\acro{LO}[LO]{local oscillator}
    \acro{NL}[NL]{nonlinearities}
    \acro{SNR}[SNR]{signal-to-noise ratio}
    \acro{SotA}[SotA]{state-of-the-art}
    \acro{TRx}[TRx]{transceiver}
\end{acronym}

\title{A Temporal Gaussian Noise Model for\\Equalization-enhanced Phase Noise}%

\author{
	Benedikt Geiger\textsuperscript{(1)}, Fred Buchali\textsuperscript{(2)}, Vahid Aref\textsuperscript{(2)}, and Laurent Schmalen\textsuperscript{(1)}
}

\maketitle                  %

\begin{strip}
    \begin{author_descr}
    
       \textsuperscript{(1)} Communications Engineering Lab (CEL), Karlsruhe Institute of Technology (KIT), \\ \hspace*{2.25ex}Hertzstraße 16, 76187 Karlsruhe, Germany,
       \textcolor{blue}{\uline{benedikt.geiger@kit.edu}}
    
       \textsuperscript{(2)} Nokia, Magirusstr. 8, 70469 Stuttgart, Germany
    \end{author_descr}
\end{strip}

\def\e{\mathrm{e}}
\def\j{\mathrm{j}}

\renewcommand\footnotemark{}
\renewcommand\footnoterule{}

\begin{strip}
    \begin{ecoc_abstract}
\!Equalization-enhanced Phase Noise causes burst-like distortions in high symbol-rate transmission systems.\! We propose a temporal Gaussian noise model that captures these distortions by introducing a time-varying distortion power. Validated through simulations and experiments, it enables accurate and simple performance prediction for high symbol-rate transmission systems.
\textcopyright2025 The Author(s)
    \end{ecoc_abstract}
\end{strip}

\tikzset{lsblock/.style = {rectangle, thick, draw, minimum width=1.4cm, minimum height=0.8cm, rounded corners=1.6mm, font=\footnotesize},}
\newcommand{\mulrel}[2]{
	\node [draw, thick, circle, minimum size = 0.3cm, #2] (#1) {};
	\draw [] (#1.south east) -- (#1.north west);
	\draw [] (#1.south west) -- (#1.north east);	
}

\section{Introduction}
\Ac{EEPN} is emerging as a significant impairment in next-generation high symbol-rate optical communication systems~\cite{xu_system_2023,xu_study_2024}. It arises from the interplay between the \ac{LO} phase noise and digital \ac{CDC}~\cite{shieh_equalization-enhanced_2008}. As a result, \Ac{EEPN} is particularly critical in cost-sensitive applications such as 1600G ZR+ transceivers~\cite{jung_mitigating_2024} and causes burst-like \ac{SNR} degradation~\cite{martins_frequency-band_2024,xu_study_2024}.

However, \ac{SotA} models treat \ac{EEPN} as \ac{AWGN} with time-invariant variance~\cite{shieh_equalization-enhanced_2008,ye_phenomenological_2022}, neglecting its time-varying behavior. As a result, the impact of \ac{EEPN} distortions is underestimated, since burst-like distortions increase the risk of uncorrectable block errors in forward error correction~\cite{xu_system_2023}. Therefore, modeling this time-varying behavior is essential for accurate link-level performance estimation.

In this work, we introduce a temporal \ac{GN} model that captures the bursty nature of \ac{EEPN} by modeling it as \ac{AWGN} with time-varying variance. To this end, we derive a simple closed-form approximation for the instantaneous distortion power and validate the model through simulations and experiments.

\section{Temporal Gaussian Noise Model for EEPN}
\begin{figure}[!t]
    \includegraphics[width=0.48\textwidth]{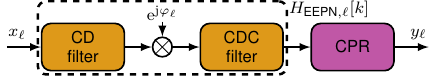}
	\caption{System model for deriving the time-varying \ac{EEPN} distortion power for the proposed temporal \acs{GN} model}
	\label{fig:system_model_derivation}
\end{figure}

We propose to model all system impairments, including \ac{EEPN}, as \textcolor{KITred}{time-varying} \ac{AWGN} since they are approximately Gaussian distributed~\cite{arnould_equalization_2019}:
\vspace*{-0.15cm}
\begin{equation}
    y_\ell \! = \! x_\ell \! \; \!+ \! \; \! n_\ell,\! \! \! \quad n_\ell \!\sim\! \mathcal{CN}(0, \sigma^2_{\text{ASE+NL+TRx}} \! \! + \! \; \! \sigma^2_{\text{EEPN},\textcolor{KITred}{\ell}}),
    \label{eq:Temporal_GN_model}
    \vspace*{-0.15cm}
\end{equation}
where $x_\ell$ and $y_\ell$ denote the transmitted and received symbols, respectively, and $\sigma^2_{\text{ASE+NL+TRx}}$ denotes the time-invariant noise power due to \ac{ASE}, fiber \ac{NL}, and \ac{TRx} impairments. The novelty of our temporal GN model is the \textcolor{KITred}{time-varying} \ac{EEPN} distortion power $\sigma^2_{\text{EEPN},\textcolor{KITred}{\ell}}$, which accounts for the burst-like nature and is derived in the sequel. Due to space constraints, we only give the key ideas and implications here. Since other distortions have a negligible interaction with \ac{EEPN}, we consider a minimal discrete-time system in which \ac{EEPN} is the only distortion, as illustrated in Fig.~\ref{fig:system_model_derivation}. The transmit symbols $x_\ell$ are passed through a \ac{CD} filter, and impaired by phase noise $\varphi_\ell$, before a \ac{CDC} filter is applied.

The \ac{CDC} filter introduces a frequency-dependent \ac{LO} phase delay~\cite{neves_enhanced_2021}, which can be approximated by an all-pass filter with frequency response~\cite{Geiger25OFC}
\vspace*{-0.15cm}
\begin{equation}
    H_{\text{EEPN},\ell}[k] \! \! = \! \! \e^{\j \varphi_{\text{EEPN},\ell}[k]}, \! \! \! \quad \varphi_{\text{EEPN},\ell}[k] \!\! = \!\! \varphi_{\ell+ R \cdot \tau_{\text{g}}[k] },
    \label{eq:phase_error_translation}
    \vspace*{-0.15cm}
\end{equation}
where $R$ is the symbol-rate, and $\tau_{\text{g}}[k]$ is the group delay introduced by the \ac{CDC} filter at frequency bin~$k$. Consequently, the phase error at each frequency bin corresponds to a delayed (or advanced) sample of the \ac{LO} phase noise, with the maximum delay at the band edges. Thus, the \ac{LO} phase noise within the \ac{CD}-induced memory ($\tau_{\text{CD}} = 2 \pi | \beta_2| LR$) is transformed into a \ac{FDPE}~\cite{Geiger25OFC}, where $\beta_2$ and $L$ are the fiber dispersion parameter and length.

The \ac{CPR} compensates the estimated phase error $\varphi_{\text{comp}} = -\bar{\varphi}_\ell$, which is the mean \ac{FDPE} in case of \ac{IDR}, as shown in \cite{MTGeiger}
\vspace*{-0.42cm}
\begin{equation}
\bar{\varphi}_\ell \approx \frac{1}{N_{\text{CD}}+1} \sum_{p = -N_{\text{CD}}/2}^{N_{\text{CD}}/2} \varphi_{\ell+p},
\label{eq:IDR}
\vspace*{-0.18cm}
\end{equation}
where ${N_{\text{CD}} = R \cdot \tau_{\text{CD}}}$ denotes the \ac{CD} memory expressed in samples. 

The \ac{EEPN} distortion power at time $\ell$ can be approximated using a sliding window of $M+1$ samples
\vspace*{-0.5cm}
\begin{align}
	\sigma^2_{\text{EEPN},\ell} & \approx \sum_{i = \ell-M/2}^{\ell+M/2} \frac{|y_i - x_i|^2}{M+1} \label{eq:derivation:ansatz}
\end{align}
\begin{align}
	\stackrel{(a)}{=} & \! \! \sum_{k = -M/2}^{M/2} \frac{\left|Y_\ell[k] - X_\ell[k] \right|^2}{M+1} \\
	\stackrel{(b)}{\approx} & \! \! \sum_{k = -M/2}^{M/2} \frac{\left| \e^{\j \varphi_{\text{comp},\ell}} \e^{\mathrm{j}\varphi_{\text{EEPN},\ell}[k]} X_\ell[k] \!- \! X_\ell[k]\right|^2}{M+1} \\
	\stackrel{(c)}{\approx} & \! \! \sum_{k = -M/2}^{M/2} \frac{\left|X_\ell[k] \right|^2 \left|\mathrm{j} \left(\varphi_{\text{comp},\ell} \! + \! \varphi_{\text{EEPN},\ell}[k] \right)\right|^2}{M+1} \\
	\stackrel{(d)}{\approx} & \! \! \sum_{k = -M/2}^{M/2} \frac{\left|\varphi_{\text{comp},\ell} + \varphi_{\text{EEPN},\ell}[k]\right|^2}{M+1} \label{eq:derivation:frequency_error} \\
    \stackrel{(e)}{\approx} & \! \! \sum_{k = -N_{\text{CD}}/2}^{N_{\text{CD}}/2} \frac{\left| \varphi_{\ell - k} - \bar{\varphi_\ell} \right|^2}{N_{\text{CD}}+1} = \mathrm{Var}_{N_{\text{CD}}+1} \left( \varphi_\ell \right)
    \label{eq:EEPN_power}
\end{align}
Here, (a) follows from Parseval's theorem with $X_\ell[k]$, $Y_\ell[k]$ as the sliding window discrete Fourier transform of $x_i, y_i, \,i\! \in\! [\ell\!-\!M/2,\ell\!+\!M/2]$; (b) applies a frequency-independent \ac{CPR}  and models \ac{EEPN} as an all-pass filter (see~(\ref{eq:phase_error_translation}) and~(\ref{eq:IDR})); (c) applies the approximation $\e^{\j a} \approx 1 + \j a$ for small phase fluctuations; (d) uses that the power spectral density of the normalized transmit symbols varies around \num{1} on a much shorter frequency-scale than the \ac{FDPE}~\cite{jung_equalization-enhanced_2025}; (e) substitutes~(\ref{eq:IDR}) and exploits that the \ac{FDPE} is given by the \ac{LO} phase noise~(\ref{eq:phase_error_translation}), thereby transitioning from the frequency to the time domain. This result shows that the \ac{EEPN} distortion power is given by the instantaneous \ac{FDPE} across $M+1$ frequency bins~(\ref{eq:derivation:frequency_error}), which, in turn, is given by the moving variance of the \ac{LO} phase noise with the channel memory length $N_{\text{CD}}+1$ as window size~(\ref{eq:EEPN_power}). %

\section{Simulation and Experimental Setup}
\begin{figure*}[!b]
	\includegraphics[width=\textwidth]{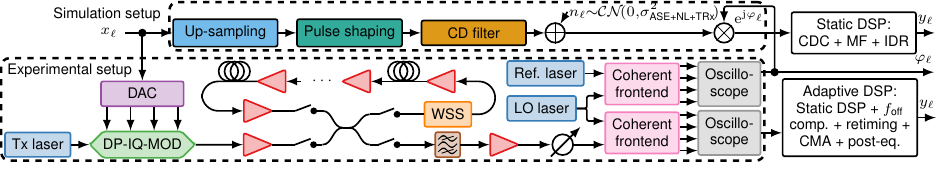}
	\vspace{-0.45cm}
	\caption{Block diagram of the experimental and simulation setup. In the experiment, we transmit a \SI{130}{GBd} signal over \SI{1900}{km} and use a second receiver to measure the \ac{LO} phase noise, which is then applied in the simulation.} %
	\label{fig:setup_block_diagram}
\end{figure*}
We verify our proposed model through both full-system simulations and experiments. The experimental setup is based on a conventional recirculating transmission loop detailed in~\cite{FredTRx,buchali_rate_2016} and includes a parallel path for measuring the \ac{LO} phase noise, as illustrated in Fig.~\ref{fig:setup_block_diagram}. At the transmitter, dual-polarized 16-QAM symbols are generated at \SI{130}{GBd} (Nyquist sampled), pre-distorted, and modulated onto a laser with a linewidth of \SI{30}{kHz}. The signal is launched into a recirculating loop comprising \num{11} fiber spans with a combined length of \SI{950}{km} and accumulated \ac{CD} of \SI{18}{ns/nm}. After $\num{2}$ round trips (\SI{1900}{km}), the signal is coherently detected at the receiver using an \ac{LO} with a linewidth of approximately \SI{210}{kHz} and sampled at \SI{256}{GSa/s}, resulting in $N_{\text{CD}} = \num{4875}$ samples at a symbol rate of \SI{130}{GBd}. To directly measure the \ac{LO} phase noise, the \ac{LO} is mixed with a low-linewidth ($<$\SI{1}{kHz}) reference laser and recorded by a second coherent receiver with a bandwidth of \SI{1}{GHz} and a sampling rate of \SI{3.125}{GSa/s}. The time-varying frequency offset between the two lasers is removed by subtracting a fifth-order polynomial approximation from the measured phase, with the residual fluctuations interpreted as the \ac{LO} phase noise. Our simulation setup, which is also shown in Fig.~\ref{fig:setup_block_diagram}, uses the same parameters as the experiment and follows~\cite{Geiger25OFC}. The transmit symbols are upsampled, pulse shaped, passed through a \ac{CD} filter, before \ac{AWGN} and \ac{LO} phase noise is added. The simulation uses the experimentally measured \ac{LO} phase noise realization. The time-invariant \ac{AWGN} noise power $\sigma^2_{\text{ASE+NL+TRx}}$ is chosen to match the experimental conditions.

In the simulations, a static \ac{DSP} chain is used, comprising resampling, \ac{CDC}, matched filtering, downsampling, and \ac{CPR} using \ac{IDR}. In contrast, the experimental setup employs an adaptive \ac{DSP} chain that additionally includes frequency offset compensation, timing recovery, polarization demultiplexing, and a post equalizer. Finally, the \ac{SNR} and distortion power are computed for blocks of $M = \num{500}$ symbols.

\section{Simulation and Experimental Results}
\begin{figure*}[!h]
        \includegraphics[width=\textwidth]{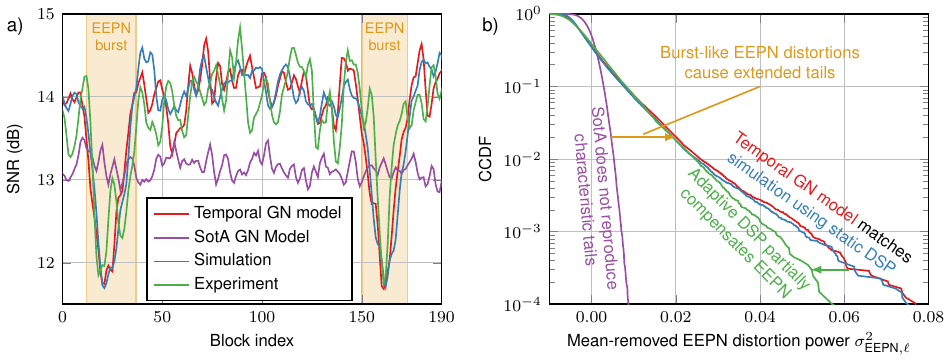}
    \vspace{-0.4cm}
	\caption{a) \ac{SNR} over time, illustrating the burst-like \ac{EEPN} in both simulation and experiment. b) Distribution of the distortion power, demonstrating \ac{EEPN} introduces extended tails. The temporal \ac{GN} model reproduces both effects, in contrast to the \ac{SotA} \ac{GN} model.}
    \label{fig:results:static_DSP}
\end{figure*}

To validate our temporal \ac{GN} model, we compare it with simulations and experiments. We benchmark it against the \ac{SotA} \ac{GN} model~\cite{shieh_equalization-enhanced_2008}, which models \ac{EEPN} as \ac{AWGN} with time-invariant distortion power. Fig.~\ref{fig:results:static_DSP}a) shows a short sequence of blockwise \acp{SNR}. We can clearly observe the burst-like degradations caused by \ac{EEPN} in simulations and experiments. While the \ac{SotA} introduces only a constant penalty, our temporal \ac{GN} model reproduces the measured \ac{SNR} fluctuations. Note that the \ac{SNR} fluctuations of the \ac{SotA} come from the small averaging length of $M = \num{500}$ symbols.

To complement the illustrative example with a statistical analysis, we show the distribution of the blockwise distortion power in Fig.~\ref{fig:results:static_DSP}b). Here, the \ac{EEPN}-induced bursts manifest as extended tails. Table~\ref{tab:correlation} gives the correlation coefficients between the measured and modeled \ac{EEPN} distortion power $\sigma^2_{\text{EEPN},\ell}$. We observe that the \ac{SotA} model fails to reproduce the tailed distribution, showing almost no correlation with the measured distortion power in Tab.~\ref{tab:correlation}. In contrast, our model reproduces the burst-induced tail distribution accurately, achieving $\rho = \num{0.93}$ with the simulation (see Tab.~\ref{tab:correlation}). Notably, our simple model remains agnostic to the \ac{LO} phase noise statistics, whereas the \ac{SotA} model relies on the assumption of a Wiener process.

\begin{figure}[!h]
    \centering
    \vspace*{-0.19cm}
    \includegraphics[width=0.48\textwidth]{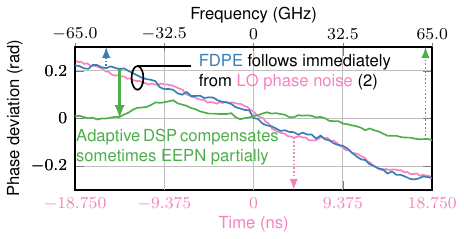}
    \vspace{-0.4cm}
    \caption{Temporal \textcolor{Set1-H}{\ac{LO} phase noise} (bottom x-axis) and frequency-dependent phase error (FDPE, top x-axis)\\for \textcolor{Set1-B}{static DSP} and \textcolor{Set1-C}{adaptive DSP}.}%
    \label{fig:phase_examples}
\end{figure}

For the experiments, we validate our model in two steps. First, we correlate the experimentally measured distortion power~(\ref{eq:derivation:ansatz}) with the \ac{FDPE}~(\ref{eq:derivation:frequency_error}), obtaining a correlation coefficient of $\rho = \num{0.95}$. This confirms that the \ac{EEPN} penalty is given by the \ac{FDPE} and validates our derivation~(\ref{eq:derivation:ansatz})-(\ref{eq:derivation:frequency_error}). Second, we observe a reduced ($\rho = \num{0.53}$) correlation with the temporal \ac{GN} model (see Tab.~\ref{tab:correlation}), because the experiments include transceiver impairments such as jitter from the signal converters, which are absent in the simulations. Additionally, the adaptive \ac{DSP} employed in the experiments includes a timing recovery, which can partially compensate \ac{EEPN}~\cite{jung_mitigating_2024}, when it manifests as a linear \ac{FDPE}, i.e., a timing offset~\cite{Geiger25OFC}. This effect is illustrated in Fig.~\ref{fig:phase_examples}, which shows the \ac{LO} phase noise and the \ac{FDPE} for both static \ac{DSP} (simulation) and adaptive \ac{DSP} (experiment) during an \ac{EEPN} burst caused by a linear \ac{FDPE}. With static \ac{DSP}, the \ac{FDPE} coincides with the temporal \ac{LO} phase noise within the \ac{CD}-induced memory, as described in~(\ref{eq:phase_error_translation}). In comparison, the adaptive \ac{DSP} reduces the \ac{FDPE}, i.e., the distortion power, leading to a shorter tail in Fig.~\ref{fig:results:static_DSP}b), and a partial decorrelation between the the \ac{FDPE}~(\ref{eq:derivation:frequency_error}) and \ac{LO} phase noise~(\ref{eq:EEPN_power}).

However, we note that only a small fraction of bursts are effectively mitigated. This is because \ac{EEPN} distortions may manifest as higher-order \ac{FDPE}~\cite{Geiger25OFC} and often vary too rapidly for the timing recovery to track. As a result, the \ac{FDPE} after adaptive \ac{DSP} mostly coincides with that of the static \ac{DSP}, matching the proposed closed-form approximation of the \ac{FDPE}~(\ref{eq:phase_error_translation}). Consequently, our simple model accurately captures the burst-like nature of \ac{EEPN}, with some bursts being partially compensated by the adaptive \ac{DSP}.

\begin{table}[!t]
    \caption{Correlation coefficient $\rho$ between the modeled and sim(ulationed)/exp(erimental) \ac{EEPN} distortion power} %
    \label{tab:correlation}
    \centering
    \setlength{\tabcolsep}{3.5pt}
    \begin{tabular}{ccc}
        \toprule
        Measured $\sigma^2_{\text{EEPN},\ell}$ & Modeled $\sigma^2_{\text{EEPN},\ell}$ & $\rho$ \\
        \midrule
        Sim. static DSP & SotA GN model & \num{0.01} \\
        Sim. static DSP & Temporal GN model & \num{0.93} \\
        Sim. adaptive DSP & Temporal GN model & \num{0.63} \\
        Exp. adaptive DSP & Temporal GN model & \num{0.53} \\
        \bottomrule
    \end{tabular}
    \vspace{-0.2cm}
\end{table}

\section{Conclusion}

We proposed a temporal \ac{GN} model that, unlike conventional approaches, captures the burst-like nature of \ac{EEPN}. To this end, we introduced a time-varying distortion power, derived as the moving variance of the \ac{LO} phase noise. We validated the model through simulations and experiments, showing that it accurately predicts \ac{EEPN}-induced impairments without any assumptions about the \ac{LO} phase noise. This makes our model a solid baseline for developing future \ac{EEPN} mitigation strategies. Finally, it enables accurate and low-complexity performance prediction for next-generation high symbol-rate transmission systems affected by \ac{EEPN}.

\clearpage
\section{Acknowledgements}
This project received funding from the European Research Council (ERC) under the European Union’s Horizon 2020 research and innovation program RENEW (grant agreement No. 101001899).

\printbibliography

\vspace{-4mm}
\end{document}